\newcounter{myctr}
\def\myitem{\refstepcounter{myctr}\bibfont\noindent\ifnum\themyctr>9\else\phantom{0}\fi\hangindent17pt\themyctr.\enskip}

\documentclass{ws-ijqi}
\usepackage{hyperref}
\usepackage[super,sort,compress]{cite}
\usepackage{url}
\usepackage{color}
\usepackage{comment} 
\usepackage{bbm}
\def\Id{{\mathbbm 1}}
\DeclareMathOperator{\Tr}{Tr}
\def\dag{{\dagger}}

\def\rrangle{{\big\rangle\!\big\rangle}}
\def\llangle{{\big\langle\! \big\langle}}
\def\RRangle{{\Big\rangle \!\! \Big\rangle}}
\def\LLangle{{\Big\langle\!\! \Big\langle}}
\def\bsigma{{\boldsymbol\sigma }}

\def\bsigmaK{\mbox{$\frac{1}{\sqrt{2}}$}\, \bsigma_k}
\def\bsigmaJ{\mbox{$\frac{1}{\sqrt{2}}$}\, \bsigma_j}
\def\bsigmaZ{\mbox{$\frac{1}{\sqrt{2}}$}\, \bsigma_0}
\def\bsigmax{\mbox{$\frac{1}{\sqrt{2}}$}\, \bsigma_1}
\def\bsigmay{\mbox{$\frac{1}{\sqrt{2}}$}\, \bsigma_2}
\def\bsigmaz{\mbox{$\frac{1}{\sqrt{2}}$}\, \bsigma_3}

\def\BF{\mathcal{E}_{\rm bf}}
\def\PhF{\mathcal{E}_{\rm phf}}

\begin{document}

\catchline{}{}{}{}{}

\title{Entanglement recovery in noisy binary quantum information protocols via three-qubit quantum error correction codes}

\author{Alessio Morea}
\address{Dipartimento di Fisica ``Aldo Pontremoli'',\\
Universit\`a degli Studi di Milano, I-20133 Milano, Italy}

\author{Michele N.~Notarnicola}
\address{Dipartimento di Fisica ``Aldo Pontremoli'',\\
Universit\`a degli Studi di Milano, I-20133 Milano, Italy\\[2ex]
INFN, Sezione di Milano, I-20133 Milano, Italy}

\author{Stefano Olivares\footnote{Corresponding author.}}
\address{Dipartimento di Fisica ``Aldo Pontremoli'',\\
Universit\`a degli Studi di Milano, I-20133 Milano, Italy\\[2ex]
INFN, Sezione di Milano, I-20133 Milano, Italy\\
stefano.olivares@fisica.unimi.it}

\maketitle

\begin{history}
\received{October 28, 2022}
\revised{\today}
\end{history}

\begin{abstract}
The task of preserving entanglement against noises is of crucial importance for both quantum communication and quantum information transfer. To this aim, quantum error correction (QEC) codes may be employed to compensate, at least partially, the detriments induced by environmental noise that can be modelled as a bit-flip or a phase-flip error channel. In this paper we investigate the effects of the simple three-qubit QEC codes to restore entanglement and nonlocality in a two-qubit system and consider two practical applications: superdense coding and quantum teleportation. Though the considered three-qubit QEC codes are known to perfectly work in the presence of very small noise, we show that they can avoid the sudden death of entanglement and improve the performance of the addressed protocols also for larger noise amplitudes.
\end{abstract}


\markboth{A.~Morea, M.~N.~Notarnicola \& S.~Olivares}
{Entanglement recovery in noisy binary quantum information protocols$\ldots$}

\section{Introduction}
Entanglement and nonlocality are genuine quantum properties of a multi-partite physical system, leading to the appearance of strong quantum correlations between its subsystems even at large distances \cite{LDent,entPurif}. As a consequence,  they have been considered so far as natural resources to perform both quantum communication and quantum information transfer \cite{E91, DenseCoding,Teleportation}. Entangled states have also been experimentally generated on different platforms such as photonic systems \cite{Aspect1, Aspect2,Aspect3,cialdi,dauria,OliRev}, trapped ions \cite{haffner2005scalable}, cavity QED \cite{PhysRevA.64.050301}, Josephson junctions \cite{bialczak2010quantum} and quantum dots \cite{gao2012observation}. Nevertheless, entanglement has been proven to be fragile against noises, which can occur in both the generation stage of states and the evolution through quantum channels \cite{Breuer,YuSD, YuEXP, PhysRevA.82.042318}. Furthermore, in the presence of mixed states it is also possible to have entanglement without nonlocality \cite{muthuganesan2018dynamics}. Given all these considerations, a crucial task is to preserve entanglement from noise.
\par
To achieve this goal, in literature there have been proposed two mainly alternative solutions: entanglement purification (EP) and quantum error correction (QEC) \cite{EP_QEC}.
EP protocols \cite{entPurif,EP1, EP2, EP3} can either detect the possible presence of errors induced by noise by performing \textit{error filtration} or \textit{entanglement distillation} by performing suitable local operations. However, these strategies requires to sacrifice some of the exploited resources, thus being limited by a rate of success.
On the contrary, QEC codes \cite{nielsenchuang, Devitt_QEC, Roffe_QEC, Lidar_QEC} implement a procedure to \textit{correct} the errors induced by noise without directly revealing their presence, through the introduction of appropriate ancillary subsystems. 
QEC codes do not require to discard any experimental run and, for this reason, they are still worth of investigation in the field of quantum communication \cite{Machiavello_QEC, Knill_QEC, Shor_QEC}.
\par
In this paper we address the performance of the application of three-qubit QEC codes to recover the entanglement of two qubits prepared in a Bell state and undergoing a noisy evolution. We consider a simplified scheme where there is a single error source, either bit flip or phase flip and, thus, we employ the associated three-qubit correcting codes \cite{nielsenchuang, Devitt_QEC, Roffe_QEC, Lidar_QEC}. Then, we characterize the obtained output state by assessing both its entanglement and nonlocality and investigate applications for two realistic protocols: superdense coding and quantum teleportation. As a matter of fact, this correction procedure is successful only in the presence of an error affecting just one of the three qubits \cite{nielsenchuang}. Therefore, the probability that an error occurs should be very small. Here we relax this request, and investigate how the performance of the QEC code is affected when the probability to have an error increases, and, thus, more than one qubit at the time can be subjected to the noise. 
\par
The structure of the paper is the following. In Sec.~\ref{sec: NC} we introduce in detail the scenario under investigation and discuss how the noisy evolution, modelled by a quantum map, affects the entanglement and nonlocality of the two qubits. Then, in Sec.~\ref{sec: C} we include QEC codes and quantify the amount of entanglement and nonlocality recovered in the system. Finally, in Sec.~\ref{sec: Appl} we discuss the superdense coding and the quantum teleportation protocols, comparing their performance with and without the application of QEC.

\section{Entanglement and nonlocality in noisy channels}\label{sec: NC}
\begin{figure}[t]
\centerline{\includegraphics[width=0.8\textwidth]{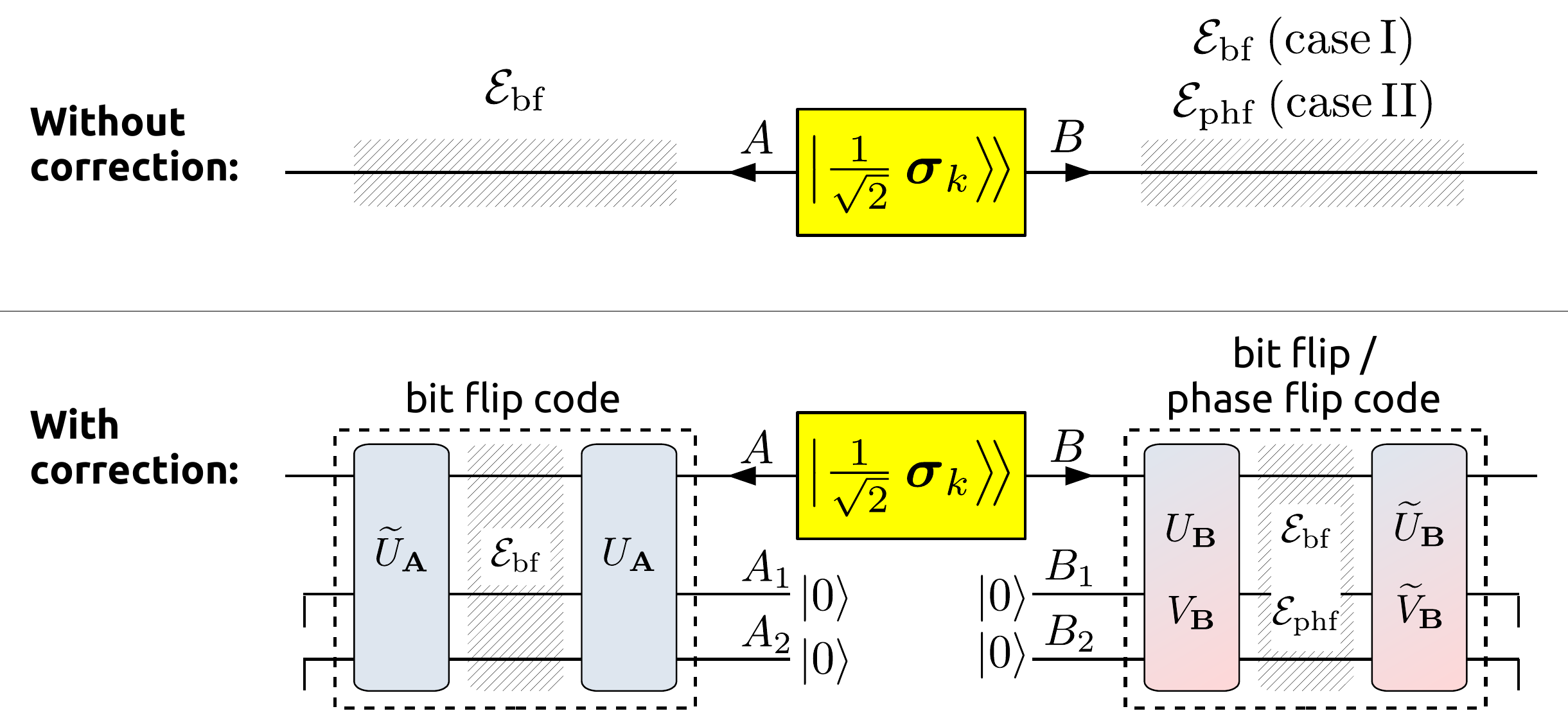}}
\caption{(Top) Two entangled qubits in a Bell state are injected into two noisy channels. In particular we address two bit-flip channels (case I) or a bit-flip channel and a phase-flip channel (case II). (Bottom) Three-qubit QEC codes are exploited to the improve the properties of the output states.}\label{fig:01-Scheme}
\end{figure}

The basic scenario discussed in this paper is depicted in the top panel of Fig.~\ref{fig:01-Scheme}. We consider a bipartite system $AB$ composed of two entangled qubits $A$ and $B$ that are injected into two noisy channels, described in terms of quantum completely-positive (CP) maps ${\cal E}$, with the net effect of degrading the entanglement of the system \cite{Breuer,YuSD, YuEXP, PhysRevA.82.042318}. We address the application of three-qubit QEC codes \cite{nielsenchuang, Devitt_QEC, Roffe_QEC, Lidar_QEC} in order to compensate, at least partially, the detriments introduced by the noisy evolution. More precisely, in this section we assess the amount of entanglement and nonlocality of the output state without QEC, whereas in the following one we employ a strategy based on QEC and compare the results obtained.
\par
The initial state of the qubits is chosen as one of the Bell states \cite{nielsenchuang}:
\begin{align}
\big| \bsigmaK \rrangle =  \frac{1}{\sqrt{2}} \sum_{l,m=0,1} 
\big[\bsigma_k \big]_{l\, m} \big | l\, m\rrangle \, , \quad (k=0,...,3) \, , 
\end{align}
where the matrix notation follows from the Choi-Jamiolkowski isomorphism \cite{CHOI, JAMIOLKOWSKI}, $| l\, m \rrangle = | l \rangle_{A} \otimes | m \rangle_{B}$ and:
\begin{align}
\bsigma_0= \begin{pmatrix} 1 & 0 \\ 0 & 1 \end{pmatrix} \, , \, \quad \bsigma_1= \begin{pmatrix} 0 & 1 \\ 1 & 0 \end{pmatrix}  \, ,\quad
\bsigma_2= \begin{pmatrix} 0 & -i \\ i & 0 \end{pmatrix} \, , \quad \bsigma_3= \begin{pmatrix} 1 & 0 \\ 0 & -1 \end{pmatrix}  \, . 
\end{align}
We note that $\llangle\frac{1}{\sqrt{2}} \bsigma_j \big | \frac{1}{\sqrt{2}} \bsigma_k \rrangle = \delta_{jk}$. 
\par
The evolution through the noisy channels affects the qubits by introducing errors with a certain probability $p$ \cite{nielsenchuang, Devitt_QEC, Roffe_QEC, Lidar_QEC}. Throughout the paper we consider two possible channels: the bit-flip and the phase-flip channels. If $\rho$ is the density matrix describing the state of a single qubit\cite{nielsenchuang}, the bit-flip channel is described by the CP map:
\begin{align}\label{eq: bitMap}
\BF (\rho) = (1-p) \, \rho + p \, \bsigma_1 \rho \, \bsigma_1 \, ,
\end{align}
which changes $|0\rangle$ to $|1\rangle$ and \textit{vice versa} with probability $p$. The phase-flip channel is described by:
\begin{align}\label{eq: phaseMap}
\PhF (\rho) = (1-p) \, \rho + p \, \bsigma_3 \rho \, \bsigma_3 \, ,
\end{align}
which instead applies a $\pi$ phase shift between $|0\rangle$ and $|1\rangle$ with probability $p$.

As emerges from Eq.s~(\ref{eq: bitMap}) and (\ref{eq: phaseMap}), the channels degrade the state, hence, in the scenario depicted in Fig.~\ref{fig:01-Scheme} (top panel), we expect a reduction of the entanglement between the two subsystems \cite{Breuer,YuSD}. 
To investigate this effect, without loss of generality we consider the input state $\big| \bsigmaZ \rrangle$ and discuss two alternative cases. In the case I we have a symmetric situation in which both qubits undergo two identical channels, i.e. both are affected by the same kind of noise. Here we address two bit-flip channels, but the same results are obtained for two phase-flip channels. In the second scenario, the case II, we are in the presence of an asymmetric setup in which the first qubit is injected in a bit-flip channel while the second in a phase-flip one (the opposite scenario would also lead to identical results).

To quantify the entanglement of the evolved states, we introduce the concurrence \cite{Concurrence1,Concurrence2} ${\cal C}$ , with $0\leq {\cal C}\leq 1$. For a generic bipartite state described by the density operator $\rho_{AB}$, it is defined as:
\begin{align}\label{eq: Concurrence}
{\cal C}= \max(0, \lambda_1-\lambda_2-\lambda_3-\lambda_4) \, ,
\end{align}
where $\lambda_1 \geq \lambda_2 \geq \lambda_3 \geq \lambda_4$ are the eigenvalues of the operator
\begin{align}
\Omega = \sqrt{\sqrt{\rho_{AB}}\, \rho_{AB}'\, \sqrt{\rho_{AB}}} \, ,
\end{align}
in which $\rho_{AB}' = \bsigma_2^{(A)} \otimes \bsigma_2^{(B)} \rho_{AB} ^* \bsigma_2^{(A)} \otimes \bsigma_2^{(B)}$ and the ``$*$'' denotes complex conjugation.
\par
Moreover, we study the quantum nonlocality by assessing the violation of the Clauser--Horne--Shimony--Holt (CHSH) inequality \cite{CHSH, CHSH2, Aspect1, Aspect2, Aspect3} using the figure of merit ${\cal B}_{\rm max}$ introduced in Ref.~\citen{HORODECKI} and defined as:
\begin{align}\label{eq: Bmax}
{\cal B}_{\rm max} = 2 \sqrt{\mu_1+\mu_2} \, ,
\end{align}
where $\mu_1 \geq \mu_2 \geq \mu_3$ are the three eigenvalues of the matrix ${\boldsymbol R}_{\rho_{AB}}= {\boldsymbol T}_{\rho_{AB}}^{\mathsf{T}} {\boldsymbol T}_{\rho_{AB}}$, with
\begin{align}\label{eq: Trho}
[{\boldsymbol T}_{\rho_{AB}}]_{n,m} = \bigg\{ \Tr\bigg[\rho_{AB} \bigg(\bsigma_n^{(A)} \otimes \bsigma_m^{(B)} \bigg)\bigg]\bigg\}_{n,m}
\end{align}
and ${n,m=1,2,3}$.
Equation~(\ref{eq: Bmax}) depends only on the two largest eigenvalues of ${\boldsymbol R}_{\rho_{AB}}$ and represents the maximum value achievable by the expectation value of the so-called Bell operator, optimized over all possible dichotomic measurements (see \ref{app:A} for details). Therefore, the CHSH inequality is violated iff ${\cal B}_{\rm max} >2$. The initial state $\big| \bsigmaZ \rrangle$ has ${\cal C}=1$ and ${\cal B}_{\rm max} =2\sqrt{2}$, that is the maximum possible values compatible with quantum mechanics laws \cite{cirelson}.

\begin{figure}[tb!]
\centerline{
\includegraphics[width=0.45\textwidth]{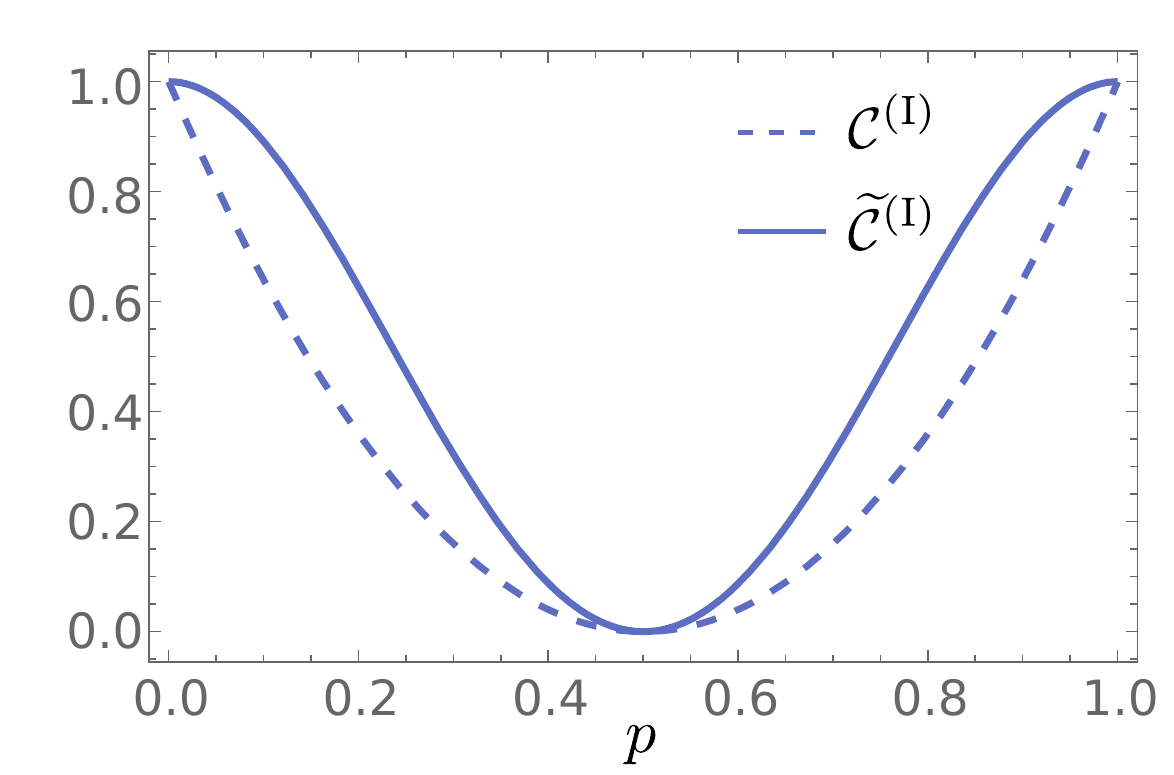} \quad
\includegraphics[width=0.45\textwidth]{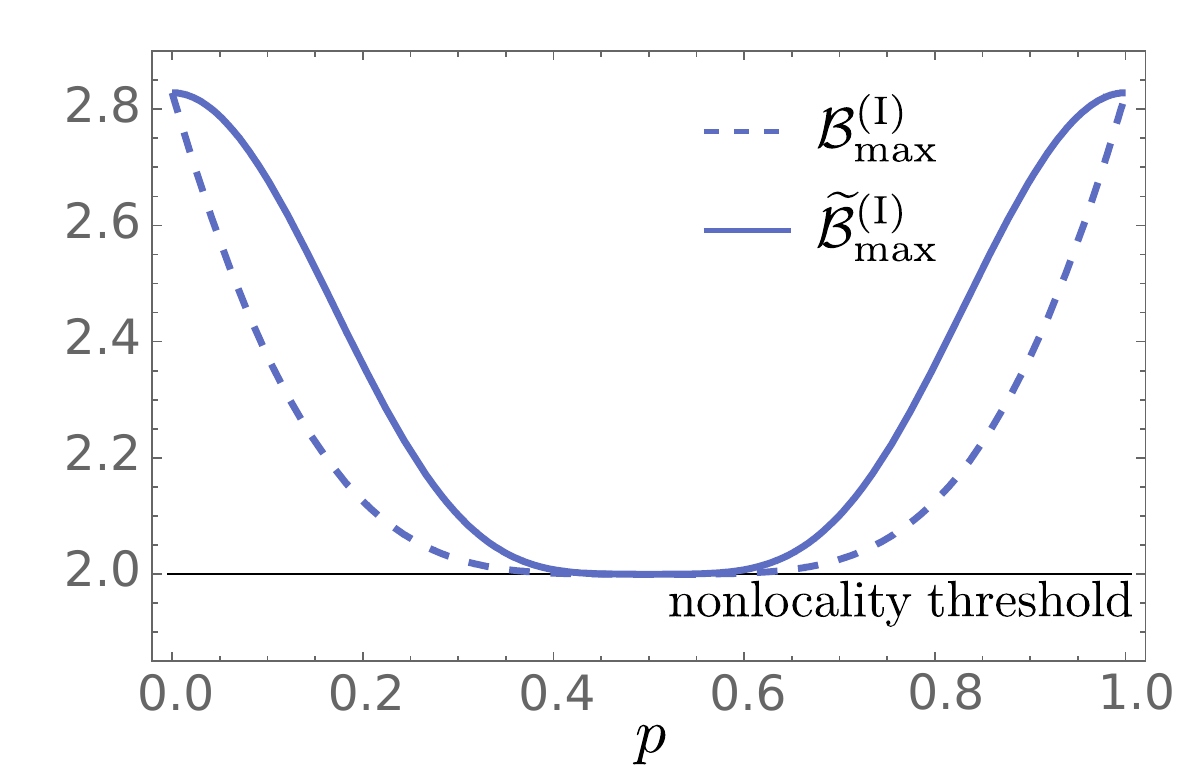} 
}
\caption{Concurrence ${\cal C}$ (left) and nonlocality ${\cal B}_{\rm max}$ (right) as functions of the channel error probability $p$ for case I (two bit-flip channels) without the application of QEC (dashed line) and with it (solid line).}\label{fig:02-BitBit}
\end{figure}

\subsection{Case I -- Two bit-flip channels}
After the evolution through the channels, the output state reads:
\begin{align}\label{eq: rho1NC}
\rho^{\rm(I)}_{AB} &=
\BF \otimes \BF \Big( \big| \bsigmaZ \rrangle \llangle \bsigmaZ \big| \Big) \, \notag \\
&= [1-2p(1-p)] \, \big| \bsigmaZ \rrangle \llangle \bsigmaZ \big| 
+ 2p(1-p) \, \big| \bsigmax \rrangle \llangle\bsigmax \big| \, .
\end{align}

The concurrence, ${\cal C}$, and nonlocality, ${\cal B}_{\rm max}$, can be computed analytically obtaining
\begin{align}\label{eq: conc NC}
{\cal C}^{\rm(I)} = (1-2p)^2  \quad \mbox{and} \quad
{\cal B}_{\rm max}^{\rm (I)} = 2 \sqrt{1+(1-2p)^4} \,,
\end{align}
respectively.
These quantities are plotted in Fig.~\ref{fig:02-BitBit}, showing that $\rho^{\rm(I)}_{AB}$ is still both partially entangled and partially nonlocal, unless for the case $p=1/2$. The choice of two identical channels implies a symmetric behaviour of ${\cal C}^{\rm(I)}$ and ${\cal B}_{\rm max}^{\rm (I)}$ with respect to $p$, since flipping both qubits allows to retrieve the original state.

\begin{figure}[tb!]
\centerline{
\includegraphics[width=0.45\textwidth]{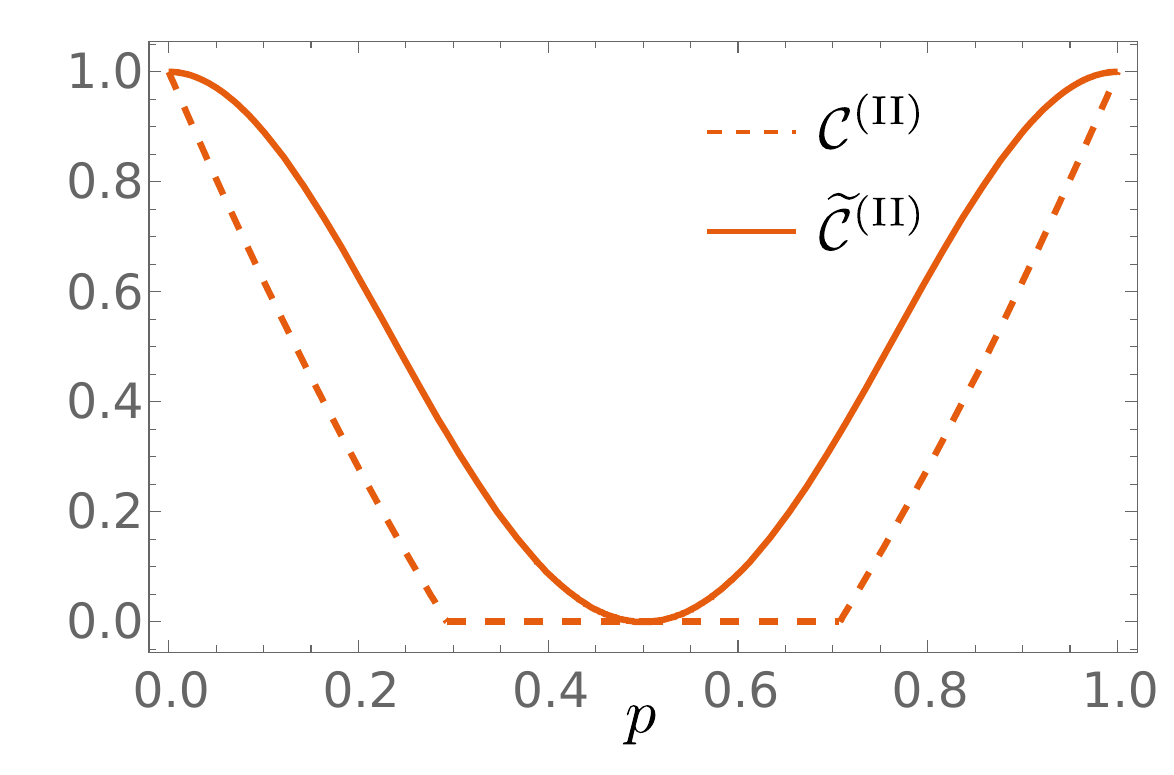} \quad
\includegraphics[width=0.45\textwidth]{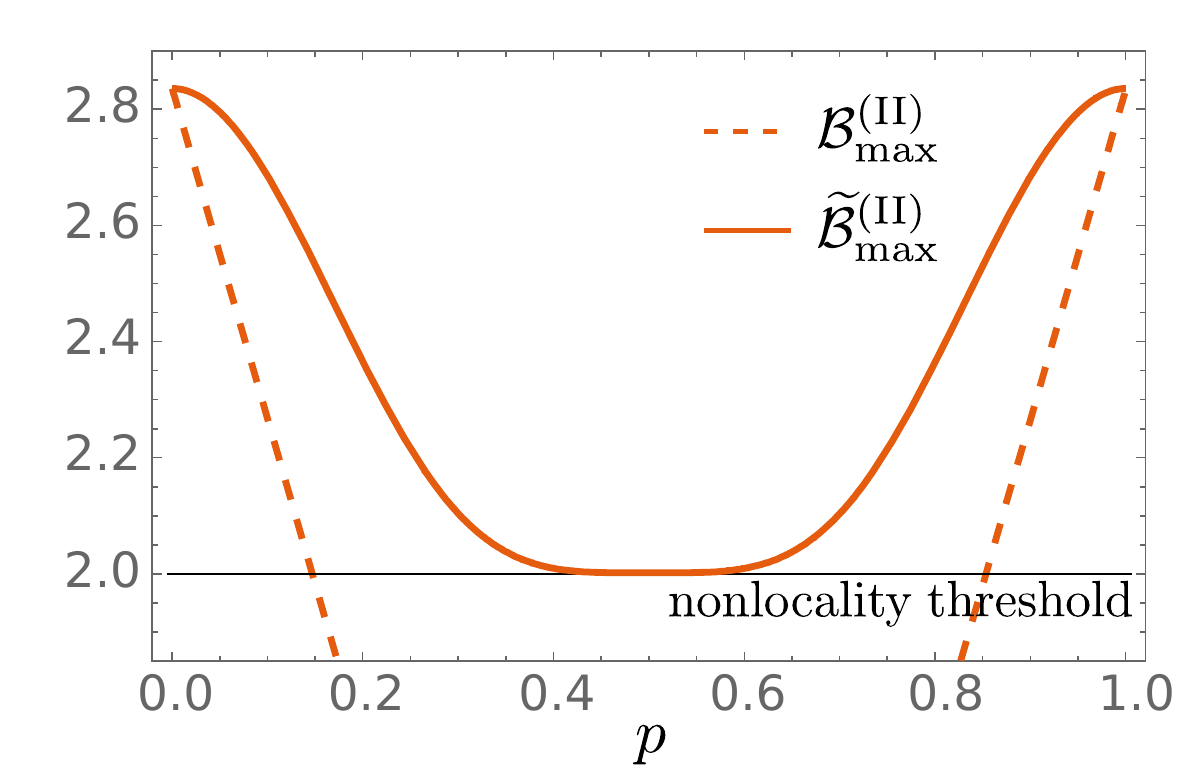}
}
\caption{Concurrence ${\cal C}$ (left) and nonlocality ${\cal B}_{\rm max}$ (right) as functions of the channel error probability $p$ for case II (bit-flip channel for qubit A and phase-flip channel for qubit B) without the application of QEC (dashed line) and with it (solid line). Employing the QEC codes prevents the sudden death of entanglement and preserves nonlocal effects for all $p \neq 1/2$.}\label{fig:03-BitPhase}
\end{figure}

\subsection{Case II -- Bit- and phase-flip channels}
Now the results are remarkably different. The final state is a mixture of all Bell states:
\begin{align}\label{eq: rho2NC}
\rho^{\rm(II)}_{AB} &=
\BF \otimes \PhF \Big( \big| \bsigmaZ \rrangle  \llangle \bsigmaZ \big| \Big) \, \notag \\
&=
(1-p)^2 \, \big| \bsigmaZ \rrangle  \llangle \bsigmaZ \big|
+ p^2 \, \big| \bsigmay \rrangle  \llangle \bsigmay \big| \, \notag \\
&\hspace{.5cm}
+ p(1-p) \,
\Big( \big| \bsigmax \rrangle  \llangle \bsigmax \big|
+ \big| \bsigmaz \rrangle  \llangle \bsigmaz \big| \Big) \, ,
\end{align}
for which the concurrence and nonlocality become
\begin{subequations}\label{eq: conc NC II}
\begin{align}
{\cal C}^{\rm(II)}= \begin{cases}
2p^2-4p+1 	&\text{if } p\leq 1-\frac{1}{\sqrt{2}} \, ,  \\
0 	&\text{if } 1-\frac{1}{\sqrt{2}}\leq p \leq \frac{1}{\sqrt{2}} \, , \\
2p^2-1 &\text{if } p \geq \frac{1}{\sqrt{2}} \, , 
\end{cases}
\end{align}
and
\begin{align}
{\cal B}_{\rm max}^{\rm (II)} = 2 \sqrt{2} |1-2p| \, ,
\end{align}
\end{subequations}
respectively, and are plotted in Fig.~\ref{fig:03-BitPhase}. Differently from case I, the concurrence vanishes abruptly at $p=p_0$, with $p_0 = (\sqrt{2} - 1)/\sqrt{2}$, leading to the sudden death of entanglement (due to symmetry reasons, for $p > 1-p_0$ the output state returns to be entangled). Moreover, for $p \in [p_0,1-p_0]$ the state is neither entangled nor nonlocal. Instead, if $p\in \big[\frac12 p_0, p_0\big]$ and
$p\in \big[1-p_0, 1 - \frac12 p_0\big]$ the state shows entanglement without nonlocality \cite{muthuganesan2018dynamics}.

\section{Preserving the states via quantum error correction}\label{sec: C}

\begin{figure}[t]
\centerline{\includegraphics[width=0.6\textwidth]{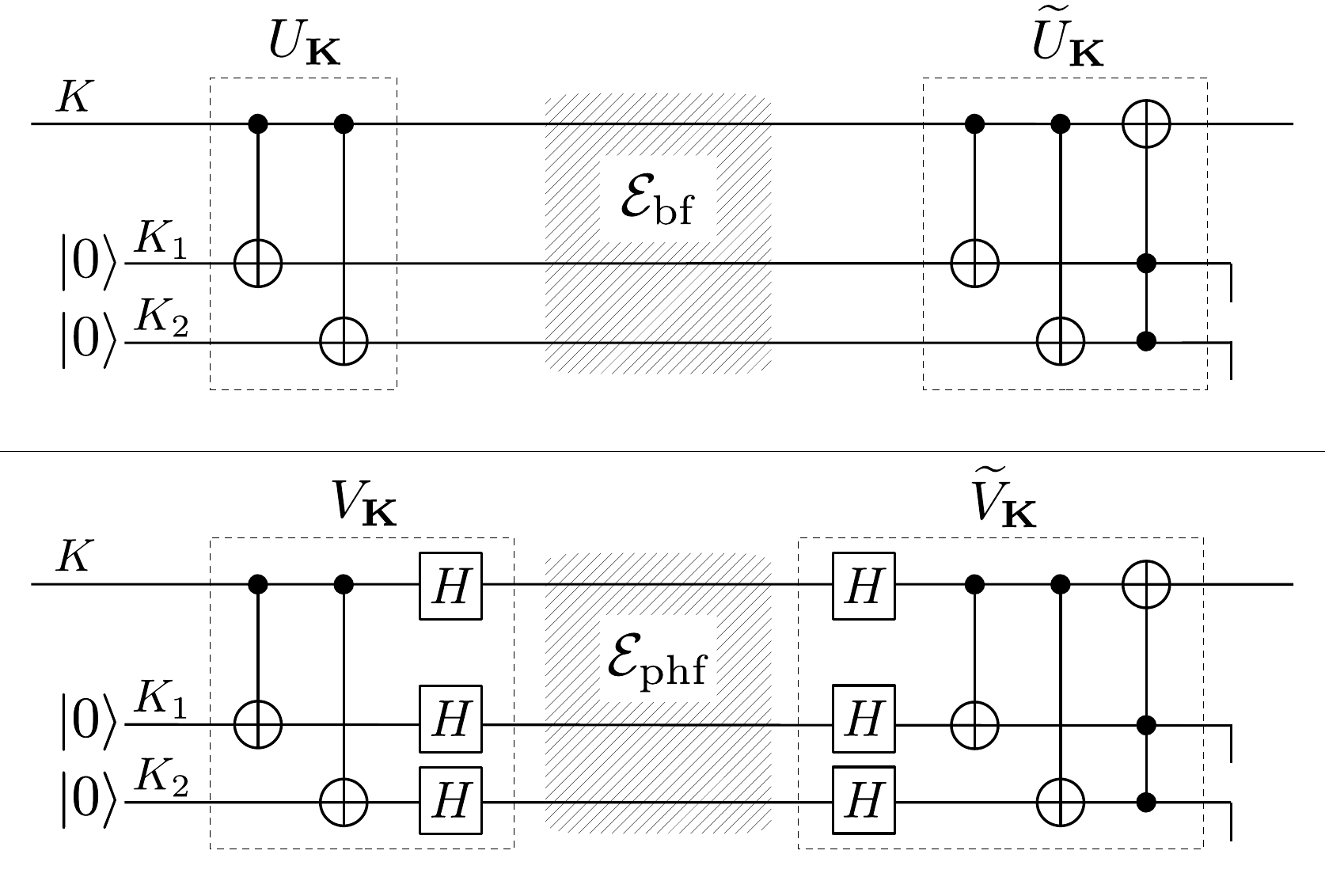}}
\caption{Schematic description of the three-qubit QEC codes acting on a single qubit $K= A,B$: the bit-flip code (top) and the phase-flip code (bottom).}\label{fig:04-Codes}
\end{figure}

In this section we investigate how using the three-qubit QEC codes \cite{nielsenchuang, Devitt_QEC, Roffe_QEC, Lidar_QEC} may be a resource to mitigate the entanglement degradation demonstrated in Sec.~\ref{sec: NC} and restore, at least partially, the lost quantum features of the state.

\subsection{Case I -- Two bit-flip channels with QEC}
The QEC code that should be applied to each of the qubits in the presence of the two bit-flip channels is depicted in the top panel of Fig.~\ref{fig:04-Codes} (see also Ref.~\citen{nielsenchuang}). It requires adding to each carrier qubit $K=A,B$ two further ancillary qubits $K_1,K_2$ prepared in state $|0\rangle_{K_1} \otimes |0\rangle_{K_2}$. A global unitary operation $U_{\bf K}$, where ${\bf K} = (K,K_1,K_2)$, consisting in two CNOT operations, is performed before sending all the three qubits into the corresponding bit-flip channels. Then, after the propagation, we apply another global unitary operation ${\widetilde U}_{\bf K}$, realized by two other CNOT operations and one Toffoli gate~\cite{nielsenchuang}. As mentioned above, only when one of the qubit is affected by the noise, or, equivalently, $p \ll 1$, $\widetilde U_{\bf K}$ leads to the full correction, and the state of the carrier qubit is completely retrieved. If $p$ increases, the induced error on the carrier qubit is eventually corrected with probability $P_{\rm corr}= (1-p)^2 (1+2p)$. 
\par
We now embed the bit-flip QEC code within the scheme of Sec.~\ref{sec: NC}. As displayed in the bottom panel of Fig.~\ref{fig:01-Scheme}, the bipartite system $AB$ of the two carrier qubits is accompanied by four ancillas as depicted in the figure. The whole system is thus prepared in state
\begin{equation}
|\Psi_{\bf AB}\rangle = | \bsigmaZ \rrangle
\otimes |0\rangle_{A_1}
\otimes |0\rangle_{A_2}
\otimes |0\rangle_{B_1}
\otimes |0\rangle_{B_2}.
\end{equation}
Before sending the qubits, we apply $U_{\bf A}$ and $U_{\bf B}$, obtaining
\begin{equation}
|\Psi^{\rm (I)}_{\bf AB} \rangle = U_{\bf A} \otimes U_{\bf B} |\Psi_{\bf AB}\rangle\,,
\end{equation}
and the subscript reminds the subsystems on which the unitary acts. Then, the state after the QEC writes:
\begin{align}\label{eq: rho1C}
\widetilde \rho^{\rm(I)}_{AB} = \Tr_{A_1 A_2 B_1 B_2} \Big\{ \widetilde U_{\bf A} \otimes \widetilde U_{\bf B}\,  {\cal N}^{\rm (I)}[\rho^{\rm (I)}_{\bf AB}]\, \widetilde U^\dag_{\bf A} \otimes \widetilde U^\dag_{\bf B} \Big\} \, ,
\end{align}
where
\begin{equation}
{\cal N}^{\rm (I)}[\rho^{\rm (I)}_{\bf AB}] =
\BF^{(A)} \otimes \BF^{(A_1)} \otimes \BF^{(A_2)}
\otimes \BF^{(B)} \otimes \BF^{(B_1)} \otimes \BF^{(B_2)}
\big(  \rho^{\rm (I)}_{\bf AB} \big)\,,
\end{equation}
namely,  each qubit is independently subjected to a bit-flip map,
and we defined
$\rho^{\rm (I)}_{\bf AB} =  |\Psi^{\rm (I)}_{\bf AB}\rangle \langle\Psi^{\rm (I)}_{\bf AB} |$.
The concurrence and the nonlocality of state~(\ref{eq: rho1C}) read:
\begin{subequations}\label{eq: conc C}
\begin{align}
\widetilde{\cal C}^{\rm(I)} = (1-6p^2+4p^3)^2 \, , 
\end{align}
\begin{align}
\widetilde{\cal B}_{\rm max}^{\rm (I)} = 2 \sqrt{1+(1-6p^2+4p^3)^4} \, ,
\end{align}
\end{subequations}
respectively, and are reported in Fig.~\ref{fig:02-BitBit}. As displayed, QEC introduces a partial mitigation of both the loss of entanglement and nonlocality. As one may expect, the mitigation is robust especially in the regime $p \ll 1$, as we can see by expanding the quantities under investigation up to the a second-order in $p$, namely, $\widetilde {\cal C}^{\rm(I)} \approx 1-12p^2$ and $\widetilde {\cal B}_{\rm max}^{\rm (I)} \approx 2\sqrt{2} (1-6p^2)$ \cite{nielsenchuang, Devitt_QEC, Roffe_QEC, Lidar_QEC}.

\subsection{Case II -- Bit- and phase-flip channels}
The QEC code for the phase-flip channel is depicted in Fig.~\ref{fig:04-Codes} (bottom panel) and it is similar to the bit-flip one. The only difference lies in the unitaries $V_{\bf K}$ and ${\widetilde V}_{\bf K}$, which, given the unitary operations $U_{\bf K}$ and $\widetilde{U}_{\bf K}$ of the bit-flip code, read $V_{\bf K}= (H_{K}\otimes  H_{K_1}\otimes  H_{K_2}) \, U_{\bf K}$ and ${\widetilde V}_{\bf K}= {\widetilde U}_{\bf K} (H_{K}\otimes  H_{K_1}\otimes  H_{K_2})$, respectively, where we introduced the Hadamard gate $H_\kappa = \frac{1}{\sqrt{2}}\big(\bsigma_1^{(\kappa)} + \bsigma_3^{(\kappa)}\big)$, acting on the mode $\kappa = K,K_1,K_2$. The correction probability is the same of the bit-flip code. 
\par
The application of the phase-flip code to the case in exam follows the same procedure as case I. At first, $|\Psi_{\bf AB}\rangle$ is transformed into:
\begin{equation}
|\Psi^{\rm (II)}_{\bf AB}\rangle = U_{\bf A} \otimes V_{\bf B} |\Psi_{\bf AB}\rangle.
\end{equation}
Then, the qubits ${\bf A}$ are sent through the bit-flip channels, while the others, ${\bf B}$, through the phase-flip ones, as depicted in the bottom panel of Fig.~\ref{fig:01-Scheme}. Finally, the ``corrected" state is:
\begin{align}\label{eq: rho2C}
{\widetilde \rho}^{\rm(II)}_{AB} =
\Tr_{A_1 A_2 B_1 B_2} \Big\{
{\widetilde U}_{\bf A} \otimes {\widetilde V}_{\bf B} \, {\cal N}^{\rm (II)}[\rho^{\rm (II)}_{\bf AB}] \,
{\widetilde U}^\dag_{\bf A} \otimes {\widetilde V}^\dag_{\bf B} \Big\} \,,
\end{align}
where, now:
\begin{equation}
{\cal N}^{\rm (II)}[\rho^{\rm (II)}_{\bf AB}] =
\BF^{(A)} \otimes \BF^{(A_1)} \otimes \BF^{(A_2)}
\otimes \PhF^{(B)} \otimes \PhF^{(B_1)} \otimes \PhF^{(B_2)}
\big( \rho^{\rm (II)}_{\bf AB} \big)\,,
\end{equation}
and $\rho^{\rm (II)}_{AB} =  |\Psi^{\rm (II)}_{\bf AB}\rangle \langle\Psi^{\rm (II)}_{\bf AB} |$.
Overall, since the Hadamard gates used in the case of the phase-flip channel change the effect of the phase flip into a bit flip\cite{nielsenchuang}, the employed QEC codes are such that ${\widetilde \rho^{\rm(II)}}_{AB} \equiv {\widetilde \rho^{\rm(I)}}_{AB}$. In turn, the concurrence and nonlocality are equal to case I:
\begin{align}\label{eq: conc C II}
{\widetilde {\cal C}}^{\rm(II)} = {\widetilde {\cal C}}^{\rm(I)} \, , \qquad
{\widetilde {\cal B}_{\rm max}}^{\rm (II)} = {\widetilde {\cal B}}_{\rm max}^{\rm (I)}  \, .
\end{align}
As a consequence, the phase-flip code prevents the sudden death of entanglement, guaranteeing entanglement and nonlocality for all $p\neq 1/2$, as displayed in Fig.~\ref{fig:03-BitPhase}.

\section{Applications to quantum information protocols}\label{sec: Appl}
\begin{figure}[tb]
\centering{\includegraphics[width=0.95\textwidth]{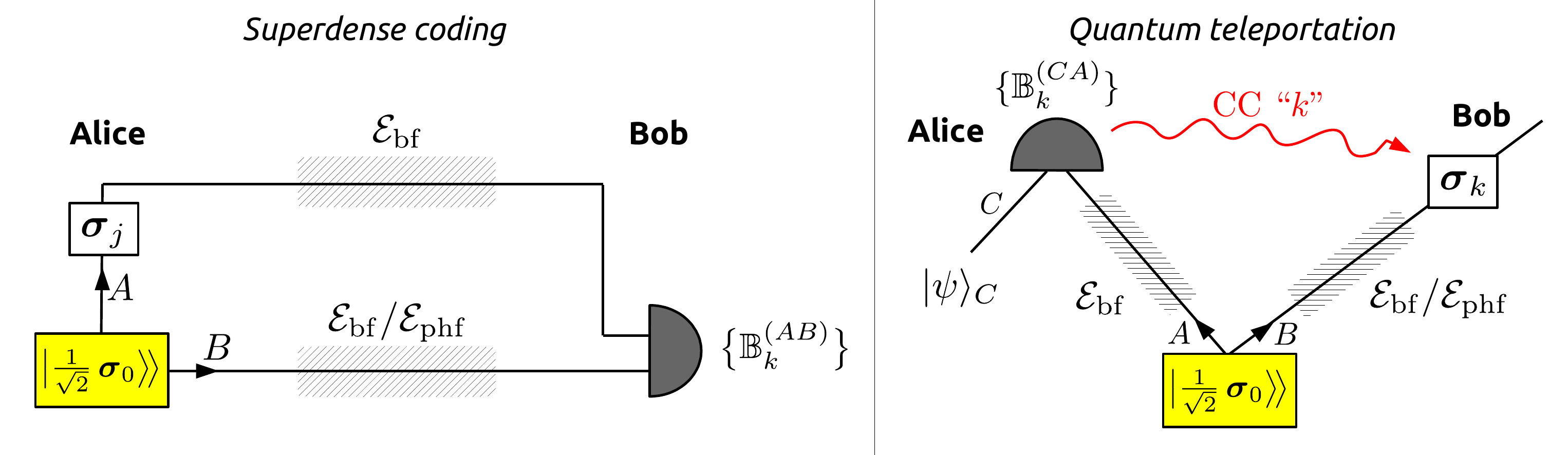}}
\caption{Scheme of the two quantum information protocols under examination: superdense coding (left) and quantum teleportation (right). The picture refers only to the scenarios without QEC.
}\label{fig:05-Appl}
\end{figure}

The capability of QEC to mitigate the detrimental effects of the noisy evolution proves it as a resource to improve the realistic performances of quantum information protocols to exchange both classical and quantum information. Here we address two paradigmatic examples: superdense coding and quantum teleportation.

\subsection{Superdense coding}
Superdense coding \cite{nielsenchuang, DenseCoding, DenseCodingExp1, DenseCodingExp2, DenseCodingExp3} is a quantum communication protocol allowing to share two bits of information by exchanging only one qubit, provided the presence of an entanglement resource. The protocol is sketched in Fig.~\ref{fig:05-Appl} (left panel). Two parties, Alice and Bob, share the Bell state $|\bsigmaZ \rrangle$ describing the qubits $A$ and $B$. Alice encodes information on her qubit via local operations, i.e. she applies $\bsigma_j$ to encode the symbol $j=0,1,2,3$. Accordingly, the entangled state shared between her and Bob is changed into $|\bsigmaJ \rrangle$. Then, she sends her qubit to Bob that, to infer the value of $j$, performs a so-called Bell measurement $\big\{\mathbb{B}^{(AB)}_k\big\}$, $k=0,1,2,3$. The measurement consist in a joint measurement on both the qubits $A$ and $B$ and corresponds to the projection onto the Bell states, namely:
\begin{align}\label{eq: BellMeas}
\mathbb{B}^{(AB)}_k= \Big| \bsigmaK \RRangle \LLangle \bsigmaK \Big| \, .
\end{align}
\par
Here we assume Alice to hold the entanglement resource and to encode the state $|\bsigmaJ \rrangle$ with equal a priori probability $q_j=1/4$ before injecting both qubits into the corresponding noisy channels. In particular, we contemplate the two cases I and II analyzed in the previous section. If QEC is not employed, the states received by Bob are 
\begin{subequations}\label{eq: DC NC}
\begin{align}
\rho^{\rm(I)}_{AB} (j)
&= \BF \otimes \BF
\Big( \Big| \bsigmaJ \RRangle \LLangle \bsigmaJ \Big| \Big) \, , \\[1ex]
\rho^{\rm(II)}_{AB} (j)
&= \BF \otimes \PhF \Big( \Big| \bsigmaJ \RRangle \LLangle \bsigmaJ \Big| \Big) \, ,
\end{align}
\end{subequations}
respectively.
\par
\begin{figure}[tb]
\centering{
\includegraphics[width=0.45\textwidth]{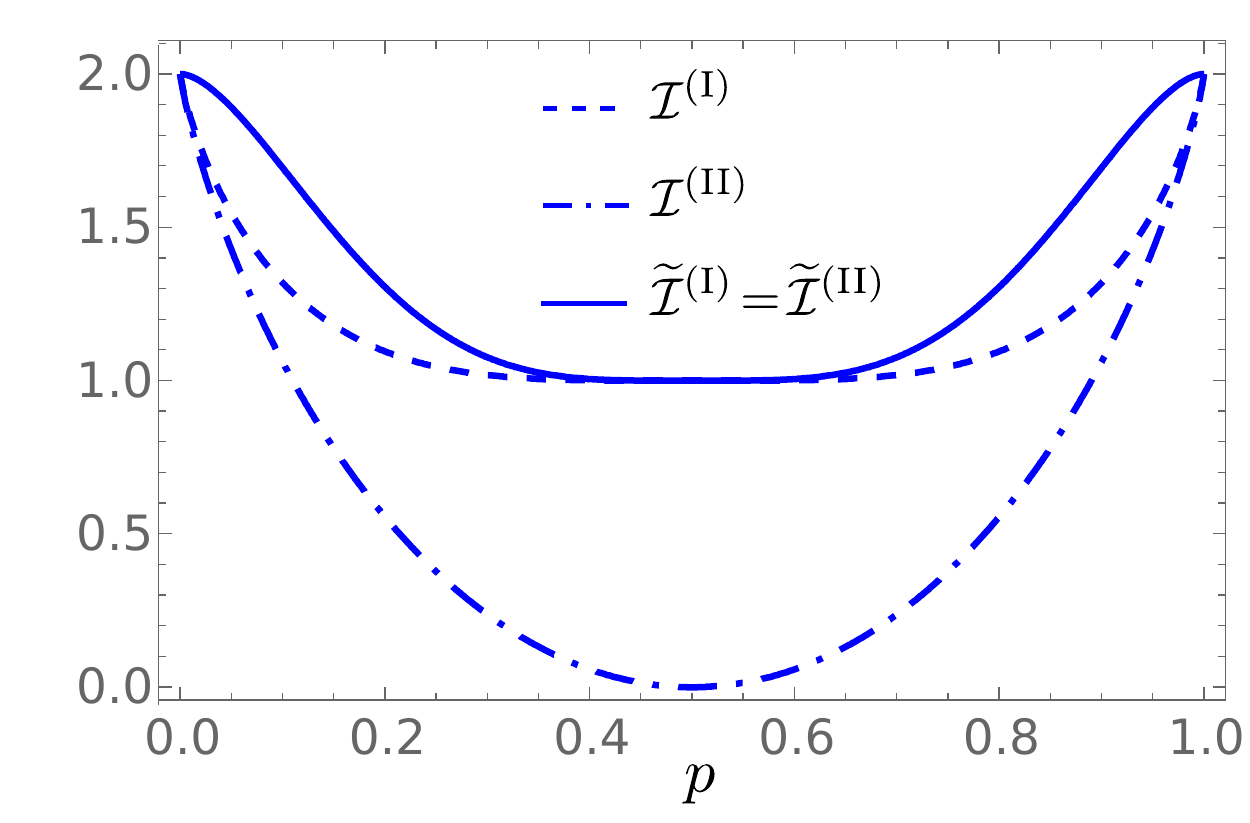}
}
\caption{Mutual information for superdense coding as a function of the channel error probability $p$ for cases I (two bit-flip channels) and II (bit-flip channel for qubit A and phase-flip channel for qubit B) without the application of QEC (dashed line: case I; dash-dotted line: case II) and with it (solid line).}\label{fig:06-Performances}
\end{figure}

Accordingly,
the conditional probability of Bob to get the outcome $k$ given $j$ reads:
\begin{align}
P^{\rm(s)} (k|j) = \Tr\bigg[ \rho^{\rm(s)}_{AB} (j) \, \mathbb{B}_k^{(AB)} \bigg] \,, \quad\mathrm{(s= I,II)} \, ,
\end{align}
and the corresponding mutual information \cite{nielsenchuang} may be written as:
\begin{align}\label{eq: MI}
{\cal I}^{\rm(s)} = \sum_{j,k=0}^3 q_j  P^{\rm(s)} (k|j) \log_2 \Bigg[ \frac{P^{\rm(s)} (k|j)}{\sum_{m=0}^3 q_m P^{\rm(s)} (k|m)} \Bigg] \,,
\end{align}
which is plotted in Fig.~\ref{fig:06-Performances} as a function of the channel parameter $p$. In the absence of QEC the mutual information decreases from 2 and reaches a minimum value equal to 1 for case I  and 0 for case II. This difference may be attributed to the symmetry induced by the exploitation of identical channels, which guarantees to preserve at least one bit of information.

By applying the three-qubit QEC procedure illustrated in the previous section, we can straightforwardly calculate the ``corrected'' states received by Bob, that is ${\widetilde \rho}_{AB}^{\rm(I)}(j)$ and ${\widetilde \rho}_{AB}^{\rm(II)}(j)$, evaluate the conditional probability:
\begin{align}
{\widetilde P}^{\rm(s)} (k|j) =
\Tr\bigg[ {\widetilde \rho}^{\rm(s)}_{AB} (j) \, \mathbb{B}_k^{(AB)} \bigg] \,, \quad \mathrm{(s= I,II)} \, ,
\end{align}
and find the corresponding mutual information ${\widetilde {\cal I}}^{\rm(s)}$, obtained by replacing all the $P^{\rm(s)}$ with ${\widetilde P}^{\rm(s)}$ in Eq.~(\ref{eq: MI}).
As we can see from Fig.~\ref{fig:06-Performances}, the embedding of QEC in the protocol mitigates the decay of the mutual information for case I and clearly improves the performance for case II, being ${\widetilde \rho}^{\rm(I)}_{AB}(j)={\widetilde \rho}^{\rm(II)}_{AB}(j)$.

\subsection{Quantum teleportation}
With the term quantum teleportation \cite{nielsenchuang, Teleportation} we refer to a protocol allowing to transfer genuine quantum information (i.e. a quantum state) from a sender to a receiver by exploiting entanglement and classical communication (CC).
Teleportation have been also experimentally demonstrated for discrete- \cite{TeleportationExp1, TeleportationExp2} and continuous-variable systems\cite{CVTeleportExp} exploiting Gaussian states\cite{OliRevPhS}.

Following the scheme of Fig.~\ref{fig:05-Appl} (right panel), a quantum source generates the Bell state $|\bsigmaZ\rrangle$ that is shared between Alice (the sender) and Bob (the receiver). Furthermore, Alice has a qubit $C$, carrying the quantum information, prepared in an arbitrary unknown state
\begin{align}\label{eq: Qubit}
|\psi(\theta,\phi)\rangle_{C}= \cos\bigg( \frac{\theta}{2} \bigg) |0\rangle_{C} + e^{i \phi} \sin\bigg( \frac{\theta}{2} \bigg) |1\rangle_{C} \,,
\end{align}
$\theta \in[0,\pi], \phi \in [0,2\pi)$ she wants to teleport to Bob. To this aim, Alice and Bob implement a protocol allowing the latter to reconstruct state $|\psi(\theta,\phi)\rangle_B$ on his shared qubit. At first, Alice performs a Bell measurement $\big\{\mathbb{B}_k^{(CA)}\big\}$ [see Eq.~(\ref{eq: BellMeas})] involving the qubits $C$ and $A$, obtaining the outcome $k$. Then, she communicates the outcome to Bob via a classically-authenticated channel (this is the CC stage of the protocol), thereafter Bob applies the unitary $\bsigma_k$ on his qubit finally retrieving the state $|\psi(\theta,\phi)\rangle_B$. We remark that this strategy requires to destroy the original state on $C$, in accordance with the no-cloning theorem \cite{nielsenchuang, Cloning1, Cloning2}.
\par
Here we assume that the qubits $A$ and $B$ are sent into two noisy channels and we still consider the cases I and II of two identical bit-flip channels or a bit-flip and phase-flip channel. As usual, we compare the performance of the protocol with and without employing QEC.
\par
In the absence of QEC, if Alice obtains outcome $k$ from the Bell measurement, Bob's qubit is reduced into the state:
\begin{align}\label{B:tele}
\rho^{\rm(s)}_{B} (k) = \frac{1}{q_k} \, \Tr_{C A} \bigg[ \rho_{C} \otimes \rho^{\rm(s)}_{AB} \, \Big(\mathbb{B}_k^{(CA)} \otimes \Id_B \Big)\bigg] \, , \quad (\mbox{s= I, II})
\end{align}
where $\rho_{C} = |\psi(\theta,\phi)\rangle_{C}\langle \psi(\theta,\phi)|$, $q_k=1/4$ is the probability of obtaining $k$ and $\rho^{\rm(s)}_{AB}$ is the state in Eq.~(\ref{eq: rho1NC}) or Eq.~(\ref{eq: rho2NC}), respectively. After the CC stage, the average state at Bob's side reads:
\begin{align}\label{eq: AveState}
\rho^{\rm(s)}_{B} = \sum_{k=0}^{3} q_k \, \bsigma_k\, \rho^{\rm(s)}_{B} (k) \, \bsigma_k \, .
\end{align}

To assess the performance of the teleportation protocol, we use as a figure of merit the average teleportation fidelity
\begin{align}
\langle {\cal F}^{\rm(s)} \rangle =
\frac{1}{4\pi} \int_{-1}^1 \!\!\! d(\cos\theta)
\int_0^{2\pi} \!\!\!\! d\phi \,
{\cal F}^{\rm (s)}(\theta,\phi) \, ,
\end{align}
where ${\cal F}^{(s)}(\theta,\phi)$ is the Uhlmann fidelity \cite{Uhlmann} between the actual output state $\rho^{\rm(s)}_{B} $ and the expected teleported one, $|\psi(\theta,\phi)\rangle_B$, that in the present case simply writes
\begin{equation}
{\cal F}^{(s)}(\theta,\phi) =  {}_B\langle \psi(\theta,\phi)|  \rho^{\rm(s)}_{B} |\psi(\theta,\phi)\rangle_B \, .
\end{equation}

\begin{figure}[tb]
\centering{
\includegraphics[width=0.45\textwidth]{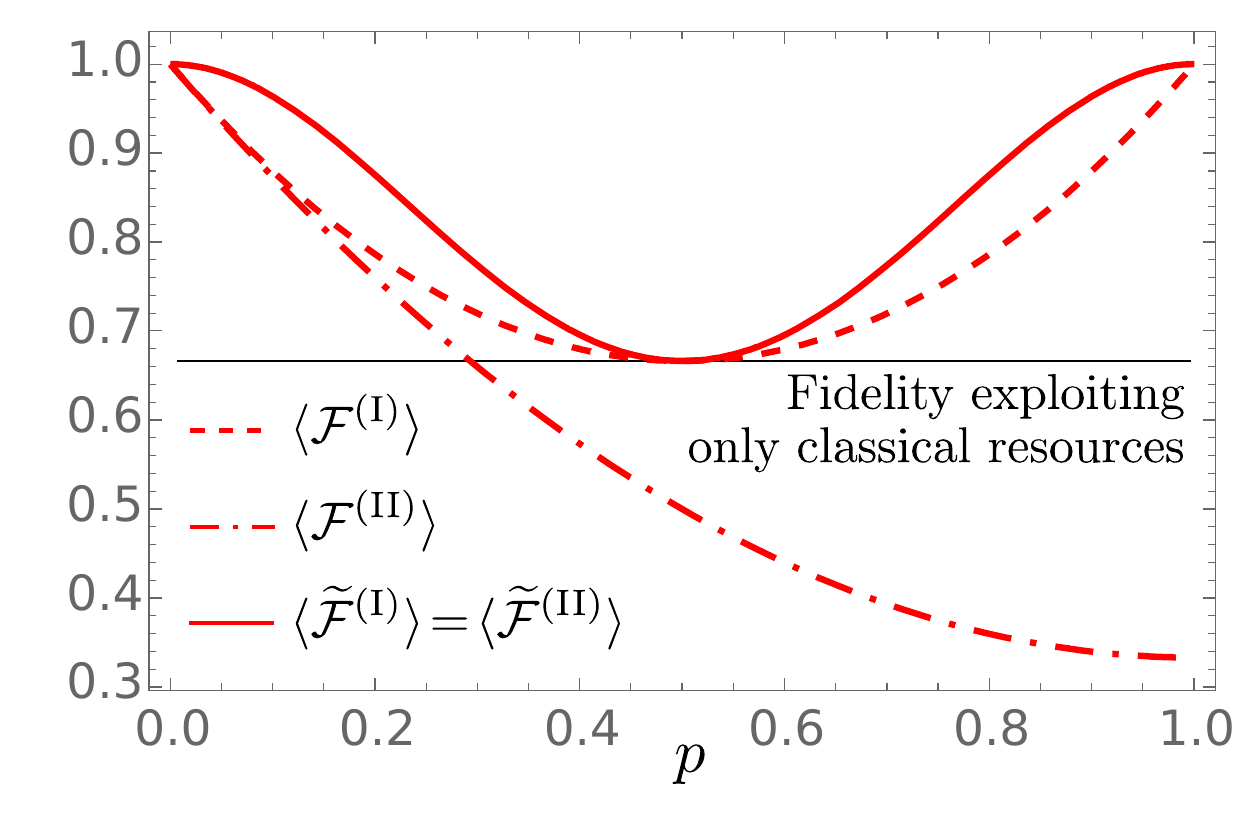}
}
\caption{Average quantum teleportation fidelity as a function of the channel error probability $p$ for cases I (two bit-flip channels) and II (bit-flip channel for qubit A and phase-flip channel for qubit B) without the application of QEC (dashed line: case I; dash-dotted line: case II) and with it (solid line). The horizontal line refers to the maximum fidelity achievable by exploiting only classical resources, namely $2/3$.}\label{fig:07-Performances}
\end{figure}

For cases I and II we obtain:
\begin{subequations}
\begin{align}
&\langle {\cal F}^{\rm(I)} \rangle = 1 - \frac43 p(1-p) \,,\\
&\langle {\cal F}^{\rm(II)} \rangle = 1 - \frac23 p(2-p) \,,
\end{align}
\end{subequations}
respectively, plotted in Fig.~\ref{fig:07-Performances}.

When QEC is performed (clearly, before Alice's Bell measurement and also before the following Bob's unitary operation) the state retrieved by Bob, $ {\widetilde \rho}^{\rm (s)}_{B}$, can be straightforwardly obtained from Eq.~(\ref{B:tele}) by replacing $\rho^{\rm(s)}_{AB}$ with the corresponding ${\widetilde \rho^{\rm(s)}}_{AB}$ given in Eq.~(\ref{eq: rho1C}) or in Eq.~(\ref{eq: rho2C}).
The average teleportation fidelity then reads:
\begin{align}
\langle {\widetilde{\cal F}}^{\rm(I)} \rangle = \langle {\widetilde{\cal F}}^{\rm(II)} \rangle =
1 - \frac43  p^2(1 - p)^2 (3 - 2 p) (1 + 2 p) \, .
\end{align}

As displayed in Fig.~\ref{fig:07-Performances}, for case I in the absence of QEC the fidelity is symmetric with respect to $p$ and if $p=1/2$ it gets its minimum value $2/3$, which coincides with the maximum fidelity achievable by using classical means, i.e. non-entangled states \cite{Cloning1}. For case II, in the absence of QEC the fidelity monotonically decreases with $p$ and reaches its minimum value $1/3$ for $p=1$. In both the cases, the introduction of QEC preserves the quantum advantage for all $p \neq 1/2$, especially in the regime $p \ll 1$ for which
$\langle {\widetilde {\cal F}}^{\rm(I)} \rangle = \langle {\widetilde {\cal F}}^{\rm(II)} \rangle \approx 1-4 p^2$.

\section{Conclusions}
In this paper we have addressed the problem of recovering entanglement in a two-qubit system undergoing a noisy evolution through bit-flip or phase-flip channels  exploiting the three-qubit QEC codes. These codes are known to perfectly restore the qubit state only in the presence of a very small noise amplitude, or, equivalently, when the probability $p$ that an error occurs is very small. Nevertheless, here we have shown that, when applied to the noisy evolution of a couple of entangled qubits, they may allow to obtain the partial recovery of both the entanglement and the nonlocality also for larger $p$. Moreover, if one of the qubits undergoes a bit-flip channel while the other a phase-flip one, QEC has been proved to prevent the sudden death of entanglement.

We have also considered the application of the ``corrected'' states to improve the performance of superdense coding and quantum teleportation in the presence of the considered noisy channels and found an improvement in the mutual information and the teleportation fidelity, respectively.

Our results, despite the simplified scenario, prove QEC as a useful resource to preserve entanglement and nonlocality in binary systems for possible applications in the field of quantum technologies in realistic conditions. 
Furthermore, they may foster more sophisticated applications, involving higher order QEC codes, e.g. Shor's nine-qubit code \cite{Devitt_QEC, PhysRevA.52.R2493} or fault-tolerant QEC techniques \cite{Devitt_QEC, Roffe_QEC, Lidar_QEC} also in the presence of other noise sources.

\section*{Acknowledgements} This work has been partially supported by MAECI, Project No.~PGR06314 ``ENYGMA'', and by University of Milan, Project No.~RV-PSR-SOE-2020-SOLIV ``S-O PhoQuLis''.

\appendix

\section{Derivation of ${\cal B}_{\rm max}$}\label{app:A}
The CHSH inequality \cite{CHSH, CHSH2} is a useful tool to verify the presence of nonlocal features in bipartite systems. It involves a scenario where a bipartite system $AB$ undergoes two distinct local measurements. On each part there can be performed two alternative measurements of dichotomic observables $\hat A_1$ or $\hat A_2$ for part $A$, and  $\hat B_1$ or $\hat B_2$ for part $B$. Then, the Bell operator is introduced as
\begin{align}\label{eq: BellOp}
\mathcal{B}= \hat A_1 \otimes (\hat B_1 + \hat B_2) + \hat A_2 \otimes (\hat B_1 - \hat B_2) \, ,
\end{align}
such that whenever its expectation value is $\langle \mathcal{B} \rangle>2$, the state of system $AB$ is nonlocal. However, $\langle \mathcal{B} \rangle$ crucially depends on the choice of the measured observables $\hat A_{1(2)}$ and $\hat B_{1(2)}$. Therefore,  for a given state we may wonder which is its maximum achievable value ${\cal B}_{\rm max}$ and consider it as a suitable figure of merit to assess nonlocal features.

For a two-qubit system, the value of ${\cal B}_{\rm max}$ may be computed analytically \cite{HORODECKI}.
Indeed, the Bell operator~(\ref{eq: BellOp}) may be rewritten as
\begin{align}
\mathcal{B}= \hat{\mathbf{a}}_1 \cdot \vec{\boldsymbol\sigma} \otimes (\hat{\mathbf{b}}_1 + \hat{\mathbf{b}}_2) \cdot \vec{\boldsymbol\sigma}  +  \hat{\mathbf{a}}_2 \cdot \vec{\boldsymbol\sigma} \otimes (\hat{\mathbf{b}}_1 - \hat{\mathbf{b}}_2) \cdot \vec{\boldsymbol\sigma} \, ,
\end{align}
where ``$\cdot$'' denotes the standard scalar product, $\vec{\bsigma}=(\bsigma_1,\bsigma_2,\bsigma_3)$ and $\hat{\mathbf{a}}_{1(2)}$ and $\hat{\mathbf{b}}_{1(2)}$ are unit vectors in $\mathbb{R}^3$ which completely characterize the operators $\hat A_{1(2)}$ and $\hat B_{1(2)}$, respectively.

If the qubits are in the state $\rho_{AB}$, the expectation value of the Bell operator reads:
\begin{align}
\langle \mathcal{B} \rangle&=  \mathrm{Tr}[\rho_{AB} \, \mathcal{B} ] \notag \\
&= \hat{\mathbf{a}}_1 \cdot {\boldsymbol T}_{\rho_{AB}} (\hat{\mathbf{b}}_1 + \hat{\mathbf{b}}_2) +  \hat{\mathbf{a}}_2 \cdot {\boldsymbol T}_{\rho_{AB}} (\hat{\mathbf{b}}_1 - \hat{\mathbf{b}}_2)  \, ,
\end{align}
where the matrix ${\boldsymbol T}_{\rho_{AB}}$ has been defined in Eq. ~(\ref{eq: Trho}). We now introduce two orthogonal unit vectors, $\hat{\mathbf{c}}_{1}$ and $\hat{\mathbf{c}}_{2}$, via the equations
\begin{align}
\hat{\mathbf{b}}_1+ \hat{\mathbf{b}}_2 = 2 \cos \theta \, \hat{\mathbf{c}}_1 \, ,
\quad \mbox{and} \quad
\hat{\mathbf{b}}_1- \hat{\mathbf{b}}_2 = 2 \sin \theta \, \hat{\mathbf{c}}_2 \, ,
\end{align}
satisfied for a certain angle $0\leq \theta \leq \pi/2$. Geometric considerations \cite{HORODECKI} lead to:
\begin{align}\label{eq: Maxc1c2}
{\cal B}_{\rm max} = 2 \, \max_{\substack{\hat{\mathbf{c}}_1, \hat{\mathbf{c}}_2 \\  \hat{\mathbf{c}}_1 \perp \hat{\mathbf{c}}_2}} \sqrt{\hat{\mathbf{c}}_1 \cdot {\boldsymbol R}_{\rho_{AB}} \hat{\mathbf{c}}_1 + \hat{\mathbf{c}}_2 \cdot {\boldsymbol R}_{\rho_{AB}} \hat{\mathbf{c}}_2 } \, ,
\end{align}
where ${\boldsymbol R}_{\rho_{AB}}= {\boldsymbol T}_{\rho_{AB}}^{\mathsf{T}} {\boldsymbol T}_{\rho_{AB}}$. Eq.~(\ref{eq: Maxc1c2}) is maximized iff $\hat{\mathbf{c}}_{1}$ and $\hat{\mathbf{c}}_{2}$ are the eigenstates of ${\boldsymbol R}_{\rho_{AB}}$ associated with its two largest eigenvalues $\mu_{1}$ and $\mu_{2}$, thus leading to Eq.~(\ref{eq: Bmax}).

\renewcommand\bibname{References}

\end{document}